\documentclass{PHYEAUTH}
\usepackage{graphicx}
\usepackage{amsmath}
\usepackage[hang]{subfigure}
\usepackage{amssymb}
\def\vc#1{\bf{#1}}
\def\mx#1{\boldsymbol{#1}}
\begin{document}

\begin{frontmatter}

\title{
Optimization of the extraordinary magnetoresistance in
semiconductor-metal hybrid structures for magnetic-field sensor
applications}

\author[address1]{M.~Holz\thanksref{thank1}},
\author[address2]{O.~Kronenwerth},
and
\author[address2]{D.~Grundler}

\address[address1]{Fachbereich Elektrotechnik, Universit\"at der Bundeswehr
   Hamburg, Holstenhofweg 85, 22043 Hamburg, Germany, and\\
 I. Institut f\"ur Theoretische Physik, Universit\"at Hamburg, Jungiusstrasse 9, 20355 Hamburg, Germany
}
\address[address2]{Institut f\"ur Angewandte Physik und Zentrum f\"ur
Mikrostrukturforschung, Universit\"at Hamburg, Jungiusstrasse 11,
20355 Hamburg, Germany}

\thanks[thank1]{
Corresponding author. E-mail: mholz@physnet.uni-hamburg.de.
Phone: +49-40-42838-2429, Fax: +49-40-42838-6798.}

\begin{abstract}
Semiconductor-metal hybrid structures can exhibit a very large
geometrical magnetoresistance effect, the so-called extraordinary
magnetoresistance (EMR) effect. In this paper, we analyze this
effect by means of a model based on the finite element method and
compare our results with experimental data. In particular, we
investigate the important effect of the contact resistance
$\rho_c$ between the semiconductor and the metal on the EMR effect.
Introducing a realistic $\rho_c=3.5\times 10^{-7}~\Omega{\rm cm}^2$
in our model we find that at room temperature this reduces the EMR
by $30\%$ if compared to an analysis where $\rho_c$ is not
considered.
\end{abstract}

\end{frontmatter}

%[main text]
\section{Introduction}
Recent studies showed that semiconductor-metal hybrid structures
can exhibit a very large geometrical magnetoresistance effect, the
so-called extraordinary magnetoresistance (EMR) effect.
Enhancements of the resistance as high as $750\,000~\%$ at $B=4~$T
have been observed~\cite{Solin:2000:1}. The EMR effect has also
been found in hybrid structures involving a two-dimensional
electron system (2DES) in an InAs
heterostructure~\cite{Moller:2002:1,Moller:2003:1}. The EMR effect
is of enormous technological interest, e.g., it has been pointed
out~\cite{Solin:2002:1} that read heads for magnetic recording
with an EMR sensor can reach storage densities in the range of
$1~$Tb/in$^{2}$. In this paper, we study the EMR effect in
rectangular hybrid structures using the finite element method
(FEM). It has been shown that the FEM, which is a numerical
technique for solving partial differential equations, provides a
powerful means of analyzing the EMR
effect~\cite{Moussa:2001:1,Holz:2003:1}. Here, we are particularly
interested in the effect of the specific contact resistance
$\rho_c$ of the semiconductor-metal interface on the optimization
of the EMR effect.

\section{Model and Potential Distribution}
\begin{figure}
\begin{center}
\leavevmode
 \hbox{\includegraphics[scale=.33]{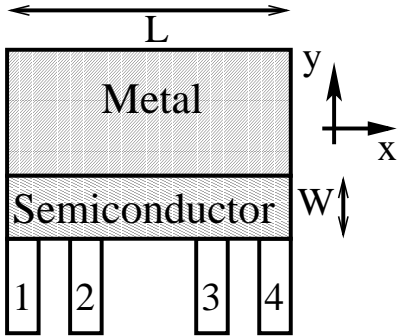}}
\end{center}
\caption{The geometry of a rectangular EMR device (top view). The contacts labelled by $1$ and $4$ are used as current contacts, while the contacts labelled by $2$ and $3$ are used as voltage probes.}
\label{fig:struktur}
\end{figure}
We consider a rectangular hybrid structure as shown in Fig.~\ref{fig:struktur} in the diffusive transport regime. The magnetic field $B$ is applied perpendicular to the $x$, $y$ plane. The current flow $\vc j$ is given by Ohm's law
 \begin{equation}
{\vc j} = {\mx \sigma} {\vc E},
\label{eq:ohm}
\end{equation}
where $\mx \sigma$ is the conductivity matrix and $\vc E$ is the electric field. All three quantities depend on the position in the hybrid structure. The conductivity matrix is given by
\begin{equation}
\mx \sigma (\beta) = \frac{\sigma_0}{1+\beta^2} \begin{pmatrix} 1 & -\beta \\ \beta & 1 \end{pmatrix}
\label{eq:matrix}
\end{equation}
with dimensionless field $\beta=\mu B$. The Drude conductivity at $B=0~\rm{T}$ is
\begin{equation}
\sigma_0=ne \mu,
\label{eq:sigma}
\end{equation}
where $n$ and $\mu$ are the carrier concentration and mobility, respectively.  By means of the continuity equation, for the steady state we obtain
\begin{equation}
{\vc \nabla} \cdot [ {\mx \sigma} \, {\vc \nabla}  \Phi (x,y) ]=0
\label{eq:poisson}
\end{equation}
for the electrical potential $\Phi(x,y)$. Inhomogeneous Neumann boundary conditions are imposed at the current contacts, while homogeneous Neumann boundary conditions are given on the rest of the device boundary. Eq.~\ref{eq:poisson} can be
solved numerically using the
FEM~\cite{Moussa:2001:1,Holz:2003:1}.
\begin{figure}
\subfigure[$B=0~$T]{
\includegraphics[scale=.5]{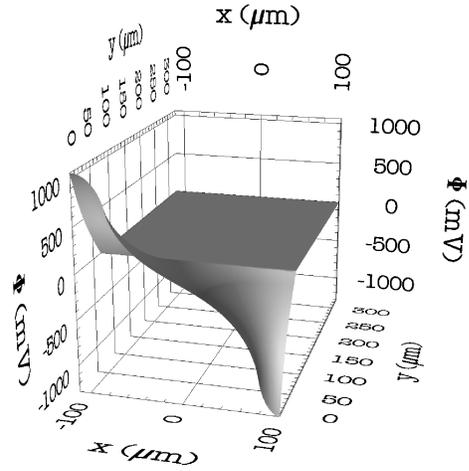}\label{fig:pot_0T}}
\hfil 
\subfigure[$B=500~$mT]{
\includegraphics[scale=0.5]{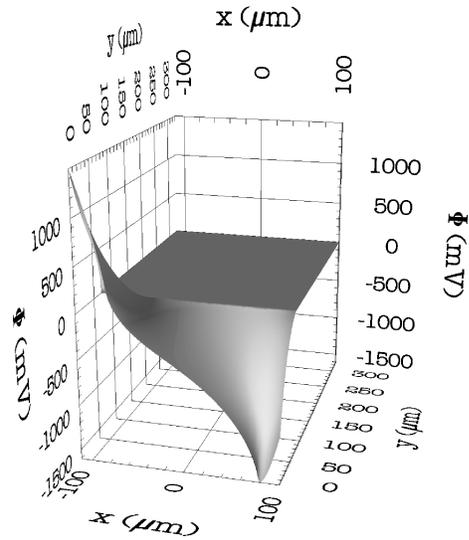}\label{fig:pot_500mT}}
\hfil \caption{ The potential distribution $\Phi(x,y)$ in an InAs(2DES)/Au
hybrid structure with dimensions $W=70~\mu$m and $L=200~\mu$m for
(a) $B=0~$T and (b) $B=500~$mT. The semiconductor is to the front,
the metal to the back of the figure. One observes a pronounced
slope of the potential at the places of the current contacts.
Here, we assumed $\mu=2.09~$T$^{-1}$, $n=3.5\times 10
^{11}~$cm$^{-2}$, and $\rho_{\rm{Au}}=2.2 \times 10^{-8}~\Omega$m.
The values of the potential are normalized to a current of $1~$mA,
so that the potential difference in mV between two points yields
directly the resistance in $\Omega$ measured between them.} \label{fig:pot}
\end{figure} In particular, the effect of
a magnetic field $B$ on $\Phi(x,y)$ can be studied in detail. An
example is given in Fig.~\ref{fig:pot}. Numerical results for the
potential distribution from Eq.~\ref{eq:poisson} are shown for a
magnetic field of $B=0~$ T [Fig.~\ref{fig:pot_0T}] and $B=500~$mT [Fig.~\ref{fig:pot_500mT}]. One
observes a pronounced voltage drop over the current contacts. The
metal film, of course, is an equipotential surface. At a magnetic
field of $500~$mT the maximum potential in
Fig.~\ref{fig:pot_500mT} is increased. Also, the potential distribution
becomes more asymmetric.

\section{Effect of Contact Resistance on Magnetoresistance}
In the following, we give results based on an InAs(2DES)/Au hybrid structure 
 studied experimentally by M\"oller {\it et al.}~\cite{Moller:2002:1} with $L=200~\mu$m and
$W=70~\mu$m. At room temperature, the carrier concentration in the
2DES is $n=3.5\times 10 ^{11}~$cm$^{-2}$ and the carrier mobility is
$\mu=2.09~$T$^{-1}$. The resistivity of the metal is
$\rho_{\rm{Au}}=2.2 \times 10^{-8}~\Omega$m. Potential
distributions in such a device have been shown in
Fig.~\ref{fig:pot}.
\begin{figure}
\begin{center}
\leavevmode
 \hbox{\includegraphics[scale=1]{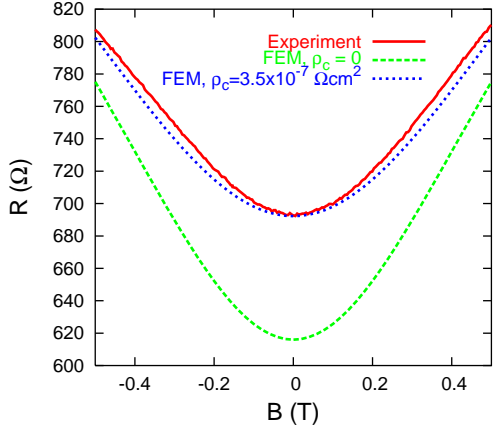}}
\end{center}
\caption{The resistance between the voltage probes $2$ and $3$
as a function of magnetic field. Here,
the case of zero contact resistance (lower broken line) is
compared to the case of $\rho_c=3.5 \times 10^{-7}~\Omega$cm$^2$ (upper broken
line). The experimental curve (full line) was taken at room temperature.}
\label{fig:R_B}
\end{figure}
In Fig.~\ref{fig:R_B}, the resistance $R$, i.e., the voltage between the contacts $2$ and $3$ divided by the current applied through the contacts $1$ and $4$, is shown as a function of magnetic field.
The cases of, both, zero contact resistance and of contact
resistance $\rho_c=3.5 \times 10^{-7}~\Omega$cm$^2$ (best fit
to experimental data) are shown. The experimental curve is also
included. One observes a good agreement for the experimental data
and the simulated curve with $\rho_c=3.5 \times
10^{-7}~\Omega$cm$^2$. Hence, it is essential to include $\rho_c$
in the analysis. We find that the value of $\rho_c$ at room
temperature is larger by a factor of 10 if compared to the value
obtained at $T=4.2$~K, where our model has given $\rho_c= 3.3 \times
10^{-8}~\Omega{\rm cm}^2$ for a hybrid
structure from the same fabrication process~\cite{Holz:2003:1}. A possible explanation is that at
elevated temperatures additional scattering of the electrons due
to bulk and interface phonons reduces the conductivity of the
semiconductor-metal interface. For technological purposes, the
value of $\rho_c$ at room temperature is most important. It is
found to be higher than at low $T$. This result suggests that the
contact resistance plays a crucial role in the optimization of EMR
devices at room temperature and should be included in the
modeling. This is done in the following, when we investigate the
interplay of the mobility $\mu$ with $\rho_c$ for low and high
magnetic fields, i.e., for $B=50~$mT and $B=1~$T, respectively.

\section{Effect of Mobility on Magnetoresistance}
%\begin{figure}
%\begin{center}
%\leavevmode
% \hbox{\includegraphics[scale=1]{./daten/R_B_70um_RT_mu_sweep.eps}}
%\end{center}
%\caption{The enhancement of $MR$ at $B=50~$mT and $B=1~$T,
%respectively. The curves for $\rho_c=0$ (dark lines) and
%$\rho_c=3.5 \times 10^{-7}~\Omega$cm$^2$ (gray lines) are
%included.} \label{fig:mu}
%\end{figure}
\begin{figure}
\subfigure[$B=50~$mT]{
\includegraphics[scale=1.]{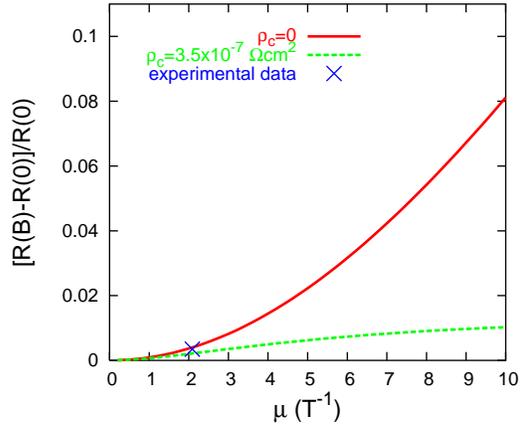}\label{fig:mu1}}
\hfil 
\subfigure[$B=1~$T]{
\includegraphics[scale=1.]{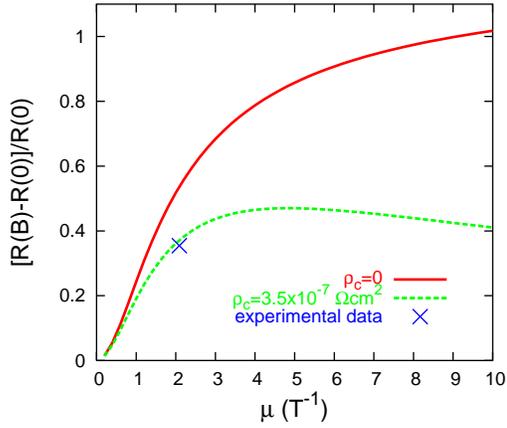}\label{fig:mu2}}
\hfil \caption{The resistance enhancement $MR$ at (a) $B=50~$mT and (b) $B=1~$T. The curves for $\rho_c=0$ (dark lines) and
$\rho_c=3.5 \times 10^{-7}~\Omega$cm$^2$ (gray lines) are
included. Please note that displayed ordinate interval in (b) is 10 times larger than in (a).}
\label{fig:mu}
\end{figure}
An increased electron mobility $\mu$ in the semiconductor affects
the  EMR behavior of the device with regard to two aspects. First,
the zero-field conductivity (Eq.~\ref{eq:sigma}) increases, i.e., the zero-field resistance $R(0)$ diminishes.
Second, the dimensionless field $\beta=\mu B$ becomes larger.
Fig.~\ref{fig:mu} shows the effect of $\mu$ on the relative
resistance change $MR=[R(B)-R(0)]/R(0)$ at $B=50~$mT [Fig.~\ref{fig:mu1}] and at
$B=1~$T [Fig.~\ref{fig:mu2}]. For a field of $B=1~$T and under the impact of a contact
resistance $\rho_c=3.5 \times 10^{-7}~\Omega$cm$^2$, a maximum
occurs in $MR(\mu)$ at about $5~$T$^{-1}$ in Fig.~\ref{fig:mu2}. This entire data set is considerably
below the trace of $\rho_c=0$. A large value of $\rho_c$ thus
limits the sensitivity of EMR devices. In Figs.~\ref{fig:mu}(a) and (b), this effect is already prominent at $\mu=2.09~$T$^{-1}$,
which is the experimentally observed value for the device of
Fig.~\ref{fig:R_B}. In Fig.~\ref{fig:mu2} the experimental $MR$
(gray cross) is smaller by about $30\%$ if compared to the case of
$\rho_c=0$. Increasing $\mu$ to $5~$T$^{-1}$ would restore
only part of the reduction. To obtain the maximum $MR$ in
Figs.~\ref{fig:mu}(a) and (b), i.e., to provide an optimized magnetic-field
sensor, a high mobility $\mu$ and, simultaneously, a low specific
contact resistance $\rho_c$ are required. This is true for the low and the
high magnetic field regime.

\section{Conclusions}
We have shown that the EMR effect can be modeled using the FEM.
In particular, the potential distribution for a realistic device
was calculated. We analyzed the effect of contact resistance and
showed that controlling $\rho_c$ is important for the optimization
of high-sensitive EMR magnetic-field sensors.

\section{Acknowledgements}
We gratefully acknowledge continuous support of the work by
H.~G{\"o}bel, D.~Heitmann, and D.~Pfannkuche, and experimental
support by C.~H.~M\"oller. We thank the Deutsche
Forschungsgemeinschaft for financial support via SFB~508 and the
BMBF via 01BM905.

\bibliography{./EMR.bib}

\begin{thebibliography}{1}
\expandafter\ifx\csname url\endcsname\relax
  \def\url#1{\texttt{#1}}\fi
\expandafter\ifx\csname urlprefix\endcsname\relax\def\urlprefix{URL }\fi

\bibitem{Solin:2000:1}
S.~A. Solin, T.~Thio, D.~R. Hines, and J.~J. Heremans, Science 289 (2000) 1530--1533.

\bibitem{Moller:2002:1}
{C.~H.~M\"oller, O.~Kronenwerth, D.~Grundler, W.~Hansen, {Ch}.~Heyn,
  and D.~Heitmann}, Appl. Phys. Lett. 80 (2002) 3988--3990.

\bibitem{Moller:2003:1}
{C.~H.~M\"oller, D.~Grundler, O.~Kronenwerth, {Ch.}~Heyn, and D.~Heitmann},
   JOSC 16 (2003) 195.

\bibitem{Solin:2002:1}
S.~Solin, D.~R. Hines, A.~C.~H. Rowe, J.~S. Tsai, Y.~A. Pashkin, S.~J. Chung,
  N.~Goel, and M.~B. Santos,  Appl. Phys. Lett. 80 (2002)
  4012--4014.

\bibitem{Moussa:2001:1}
J.~Moussa, L.~R. Ram-Mohan, J.~Sullivan, T.~Zhou, D.~R. Hines, and S.~A. Solin,
   Phys. Rev. B 64 (2001) 184410.

\bibitem{Holz:2003:1}
{M.~Holz, O.~Kronenwerth, and D.~Grundler},  Phys. Rev. B 67 (2003) 195312.

\end{thebibliography}

\end{document}